\documentclass[aps,prl,twocolumn,showpacs]{revtex4}

\usepackage{graphicx}
\usepackage{dcolumn}
\usepackage{bm}
\usepackage{epsfig}
\usepackage{epstopdf}
\usepackage{subfigure}

\begin{document}

\title{Universal condition for critical percolation thresholds of kagom\'e-like lattices}

\author{Robert M. Ziff}
\email{rziff@umich.edu}

\author{Hang Gu}
\email{ghbright@umich.edu}

\affiliation{Michigan Center for Theoretical Physics and Department of Chemical Engineering,           University of Michigan, Ann Arbor MI 48109-2136}

\date{\today}
\begin{abstract}
Lattices that can be represented in a kagom\'e-like form are shown to satisfy a universal percolation criticality condition, expressed as a relation between $P_3$, the probability that all three vertices in the triangle connect, and $P_0$, the probability that none connect.  A  linear approximation for $P_3(P_0)$ is derived and appears to provide a rigorous upper bound for critical thresholds.  A numerically determined relation for $P_3(P_0)$ gives thresholds for the kagom\'e, site-bond honeycomb, (3-12$^2$) lattice, and ``stack-of-triangle" lattices that compare favorably with numerical results.
\end{abstract}
\maketitle

Percolation is the study of long-range connectivity in random systems.   The value of the site or bond occupation probability where that connectivity first appears is percolation threshold $p_c$  \cite{StaufferAharony94}.
Finding exact and approximate $p_c$'s for percolating systems on various lattices is a long-standing problem that continues to receive much attention today
(e.g., \cite{Kondrat08,RiordanWalters07,ScullardZiff06,ZiffScullard06,Ziff06,Scullard06,ScullardZiff08,Parviainen07,QuintanillaZiff07,NeherMeckeWagner08,JohnerGrimaldiBalbergRyser08,Ambrozic08,FengDengBlote08,Wu06,MajewskiMalarz07,WagnerBalbergKlein06,TarasevichCherkasova07,MayWierman05,HajiakbariZiff08}).  

 \begin{figure}
\includegraphics[width=0.6\hsize]{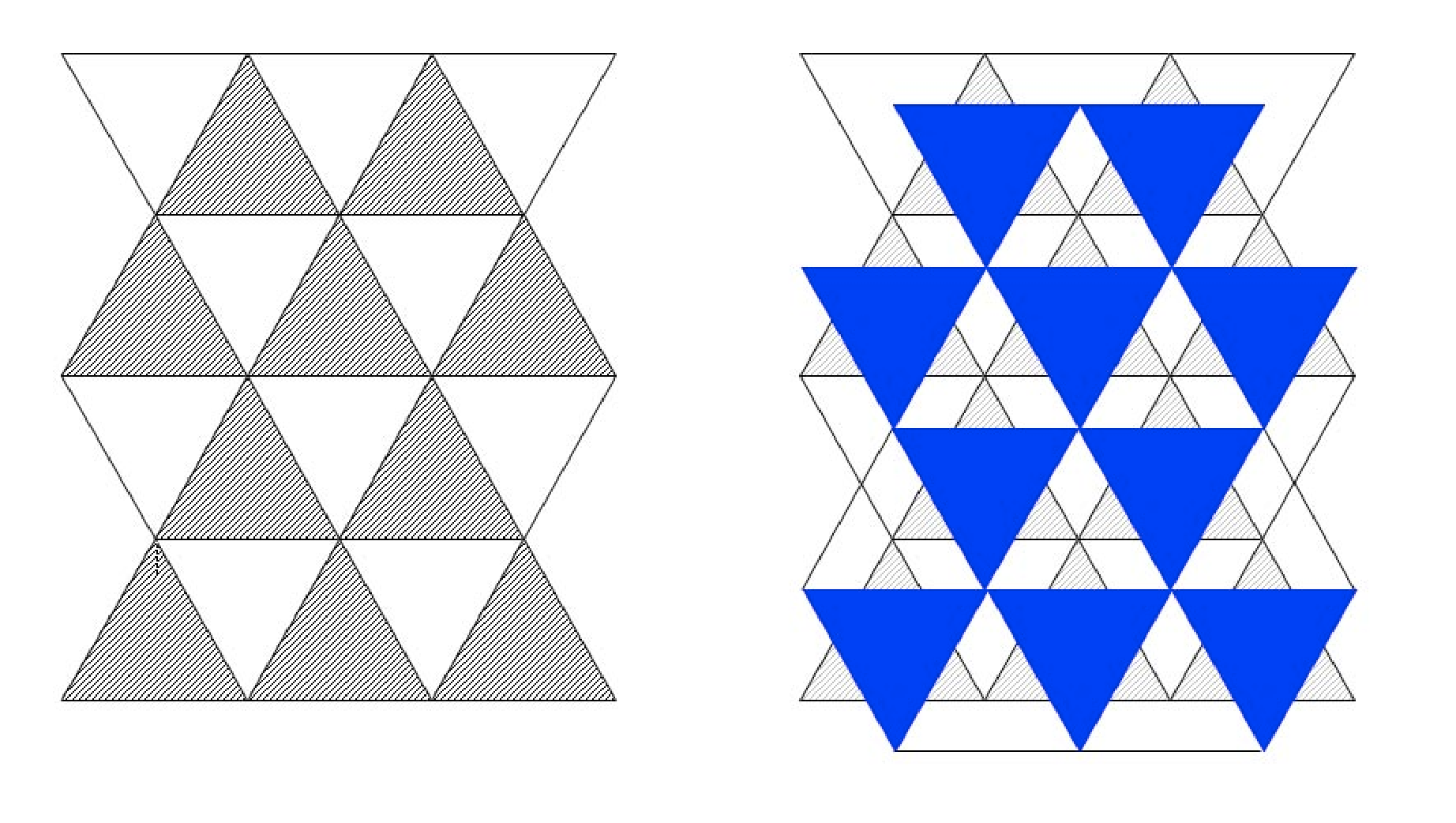}
\caption{$\triangle$-$\triangle$ duality for lattices in simple triangular array.  (left) Shaded triangles represent any collection of internal bonds. (right) Result of $\triangle$-$\triangle$ transformation where blue (dark) triangles are the dual triangles, and form the same arrangement as on the left but rotated $180^\circ$.}
\label{tritri}
\end{figure}

All known exact $p_c$'s are for two-dimensional lattices that can be represented as arrays of triangular units self-dual in the triangle-triangle ($\triangle$-$\triangle$) transformation, as illustrated in Fig.\ \ref{tritri}  for the case of a simple triangular array.  When this duality is satisfied, $p_c$ is determined by the simple condition   \cite{Ziff06,ChayesLei06}
\begin{equation}
P_3' = P_0' \ ,
\label{P30}
\end{equation}
where $P_3'$ is the probability that all three vertices of the triangular unit connect, $P_0'$ is the probability that none connect, and the prime indicates a $\triangle$-$\triangle$-dual system.   The shaded triangular units can contain any collection of bonds, including correlated bonds which can mimic site percolation, connecting the three vertices. 

If, for example, the triangular unit is simply a triangle of three bonds, each occupied with probability $p$,
then $P_0' = q^3$ and $P_3' = p^3 + 3 p^2 q$, where $q = 1 - p$, and (\ref{P30}) yields the bond criticality condition for the triangular lattice as $q^3 = p^3 + 3 p^2 q$ which has the solution $p_c = 2 \sin \pi/18 = 0.34729636$ \cite{SykesEssam64}.  Likewise, taking a star of three bonds as the basic unit gives $P_0' = q^3 + 3 q^2 p$ and $P_3' = p^3$, and (\ref{P30}) yields $ q^3 + 3 q^2 p = p^3$ or  $p_c = 1 - 2 \sin \pi/18 = 0.65270365$ for the honeycomb lattice \cite{SykesEssam64}. 
Eq.\ (\ref{P30}) has been applied to many other lattices that satisfy $\triangle$-$\triangle$ duality, including ``martini" \cite{Scullard06,Ziff06,Wu06}, bowtie \cite{Wierman84,ZiffScullard06}, and ``stack-of-triangle" \cite{HajiakbariZiff08} lattices, to find exact $p_c$'s. 

\begin{figure}
\includegraphics[width=0.5\hsize]{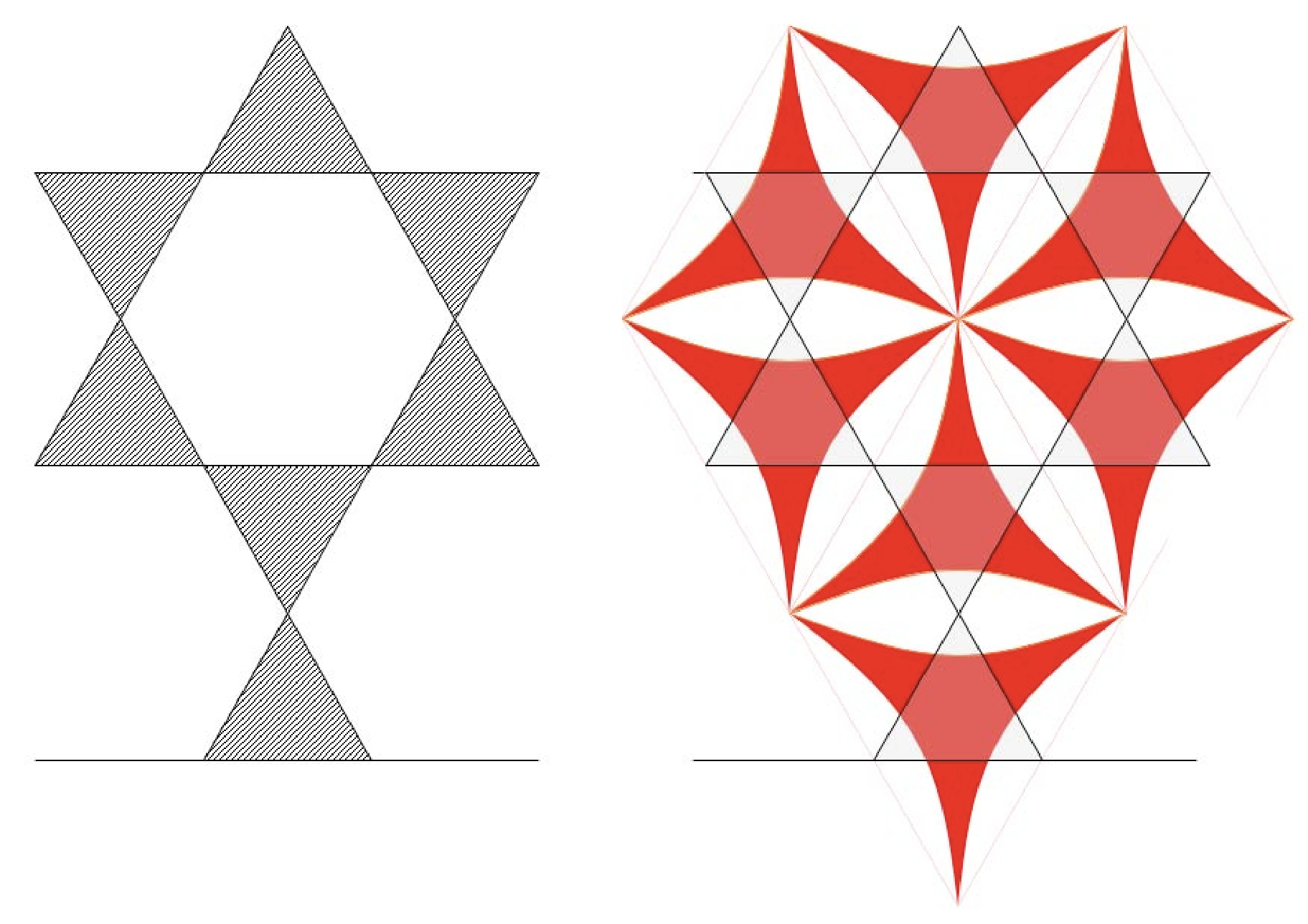}
\caption{(left) Shaded triangles in the generalized kagom\'e configuration.  (right) Result of $\triangle$-$\triangle$ transformation, showing that this system is not self dual.}
\label{kagdual}
\end{figure}

\begin{figure}
\subfigure[]{\label{honeycombfig}}
\includegraphics[width=0.35\hsize]{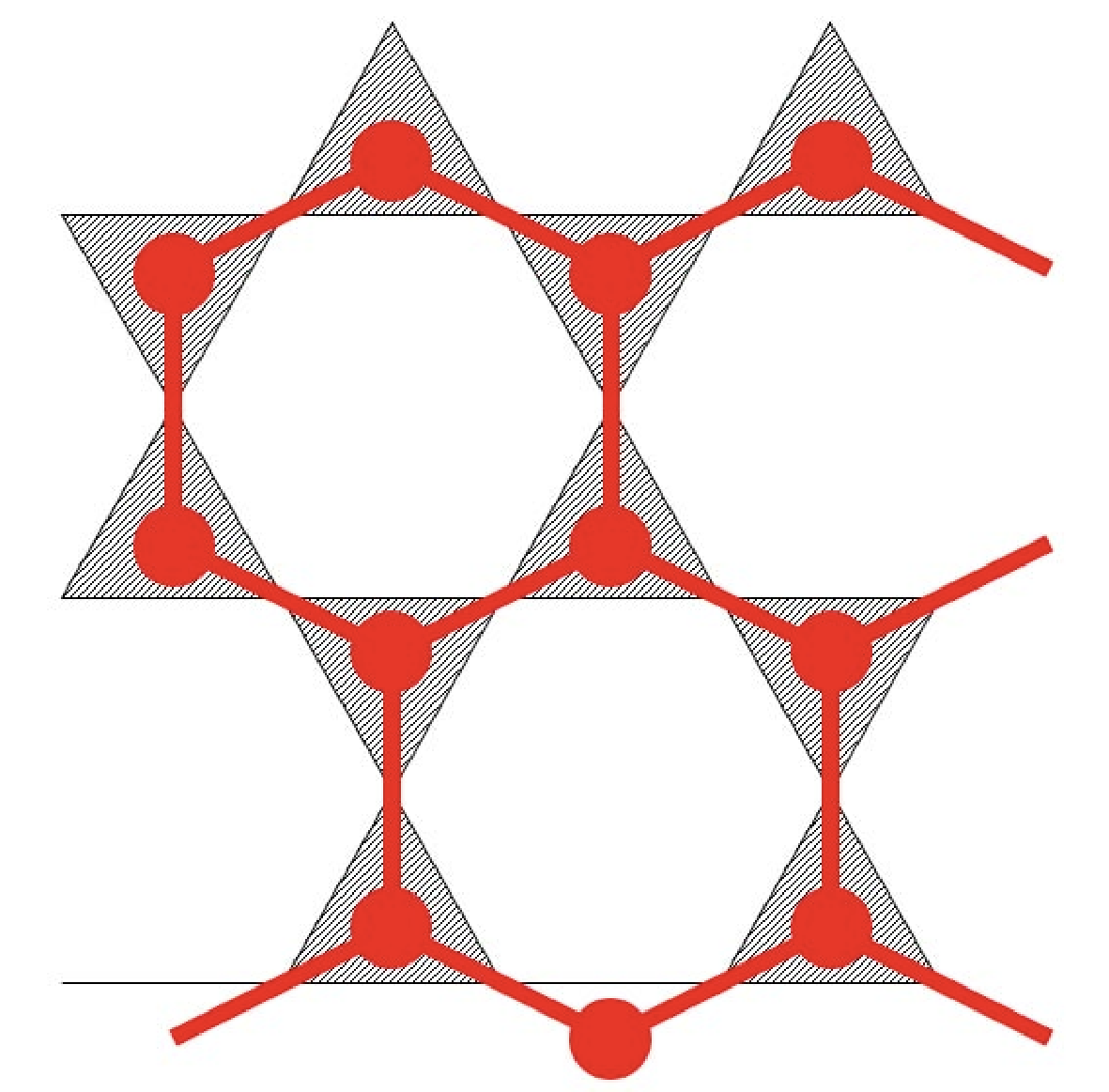}
\subfigure[]{\label{sitebondfig}} \includegraphics[width=0.35\hsize]{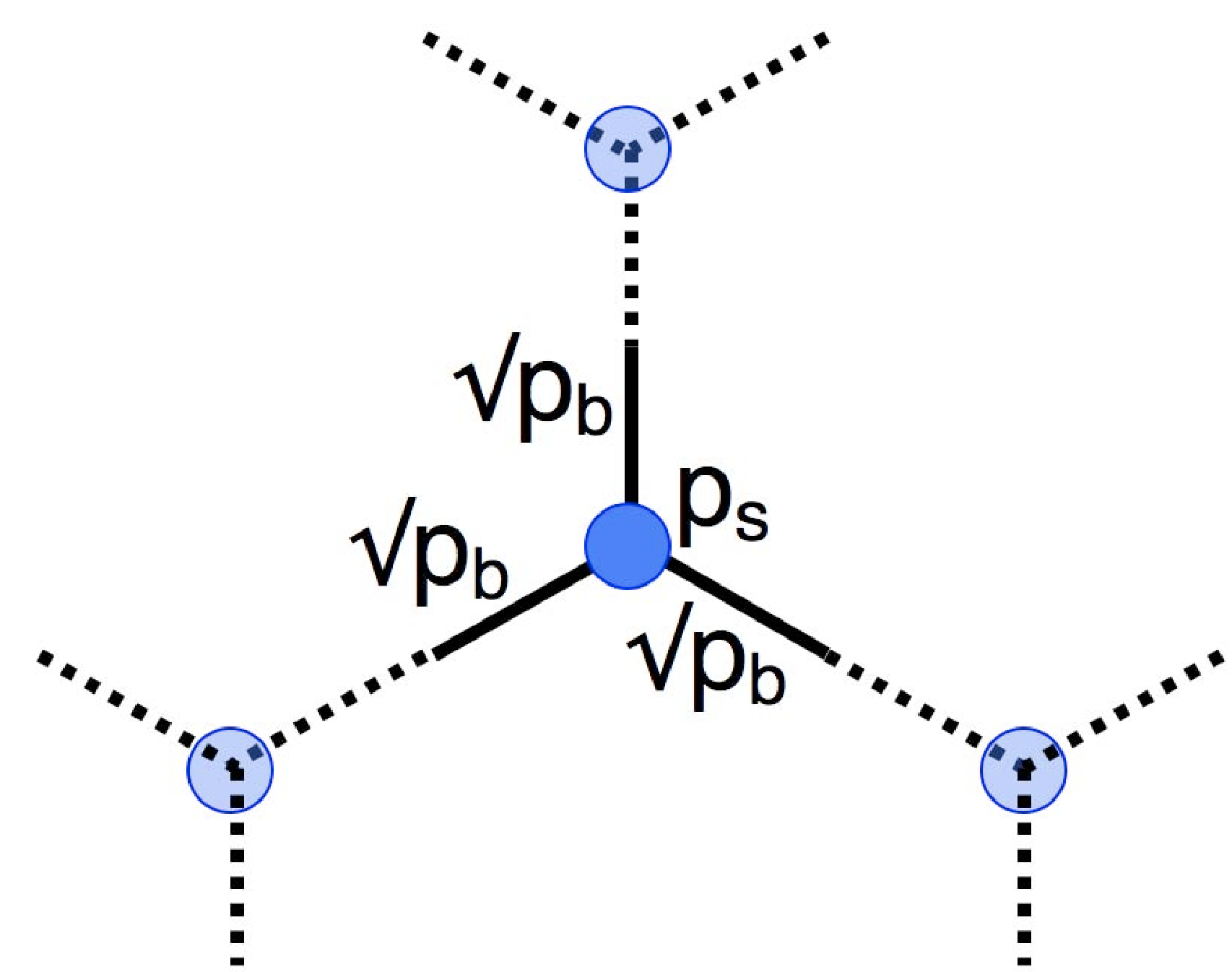}
\caption{(a) Site percolation on the honeycomb lattice (red) is equivalent to bond percolation on the kagom\'e lattice (shaded triangles) with all three bonds occupied, or all vacant.  (b) Basic unit for analyzing site-bond percolation on the honeycomb lattice in the generalized kagom\'e framework.}
\end{figure}

However, when $\triangle$-$\triangle$ duality is not satisfied, then  Eq.\ (\ref{P30}) cannot be used to find $p_c$.  For example, the $\triangle$-$\triangle$ transformation for the kagom\'e lattice is shown in Fig.\ \ref{kagdual}, and it can be seen that, while the lattice can be broken up into non-touching shaded triangular units, the $\triangle$-$\triangle$ transformation  gives a different lattice altogether, and so the self-duality condition is not satisfied.  Likewise, site percolation on the honeycomb lattice, which can be represented as bond percolation on the kagom\'e lattice with all three bonds correlated (see Fig.\ \ref{honeycombfig}), is also non-self-dual.

\begin{figure}
\includegraphics[width=0.8\hsize]{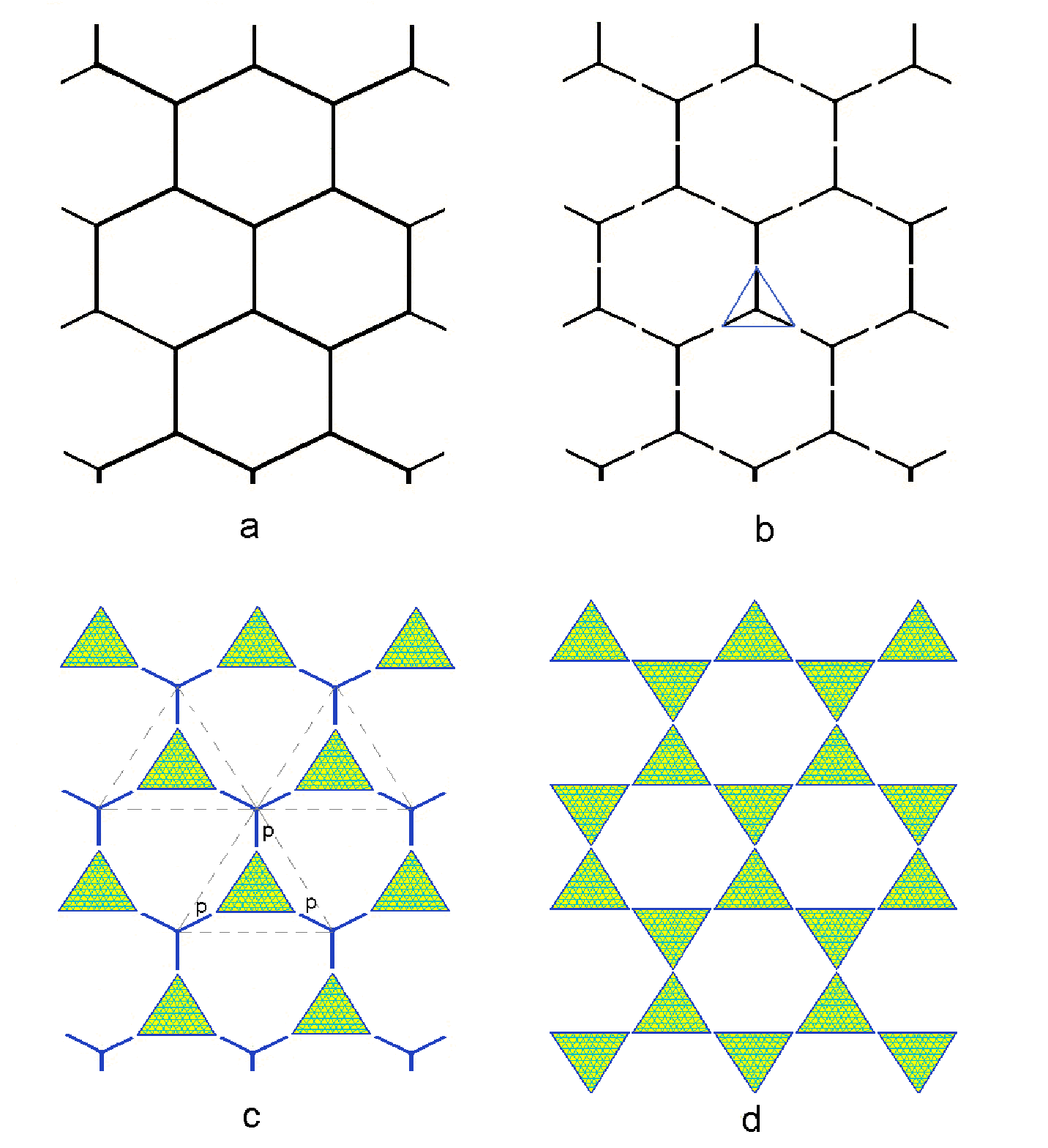}
\caption{Steps in the derivation of the linear relation Eq.\ (\ref{linear}): (a) the honeycomb lattice, (b) double-honeycomb forms a kagom\'e class of lattice, (c)  all up-stars replaced by triangular units, forming martini configuration satisfying $\triangle-\triangle$ duality,  (d) remaining stars replaced by triangular units, forming the kagom\'e configuration.}
\label{fourfigs}
\end{figure}

Nevertheless, for any system that can be broken up into identical disjoint isotropic triangular units, $p_c$ must be determined by a unique condition that depends only upon the connections probabilities $P_0$ and $P_3$ of the triangular units.  In this paper we consider lattices of the kagom\'e form, as shown in  \ref{fourfigs}(d), and investigate the corresponding relation between $P_3$ and $P_0$.  The kagom\'e form includes several unsolved lattices of interest as discussed below. 
While we can't find exact thresholds for these lattices (indeed, they are likely insolvable), we can make very precise predictions on their values and unify their study.

First we consider the ``double honeycomb" lattice, shown in Fig.\ \ref{fourfigs}(b), which is of the kagom\'e form and is the one exactly soluble lattice of this form. It can be constructed by replacing each bond of a honeycomb lattice (Fig.\ \ref{fourfigs}(a)) by two bonds in series, which implies that its $p_c$ is the square root of the $p_c$ for the honeycomb lattice:
\begin{equation}
p^\star = \sqrt{1 - 2 \sin \pi /18} = 0.80790076
\end{equation}  
For this lattice, which we indicate by a star, we have
\begin{eqnarray}
P_0^* &=& {q^\star}^3 + 3 {q^\star}^2 p^\star = 0.09652861 \label{P0star} \\
P_3^* &=&  {p^\star}^3 = 0.52731977
\label{P3star}
\end{eqnarray}
where $q^\star = 1 - p^\star$.  Note, Eq.\ (\ref{P30}) is far from being satisfied. 

Next, generalizing the considerations in \cite{ScullardZiff06}, we develop an approximate linear relation between $P_3$ and $P_0$ for all lattices of the kagom\'e form, that is exact at the point $(P_0^*$, $P_3^*)$.   Consider the  systems shown in Fig.\ \ref{fourfigs}. In (c) we replace all the up-stars of (b) with general shaded triangular units with a given net connectivity $P_0$ and $P_3$.    This produces a generalized ``martini" configuration, which falls under the general triangular class of Fig.\ \ref{tritri}, with connectivities (as follows from the diagram in (c)):
\begin{eqnarray}
P_0' &=& P_0 +3 P_2 ({q^\star}^2 + 2 q^\star p^\star) + P_3 ({q^\star}^3 + 3 {q^\star}^2 p^\star) \nonumber \\
P_3' &=& P_3 {p^\star}^3 
\end{eqnarray}
Eq.\ (\ref{P30}) then yields the exact criticality condition for system (c):
\begin{equation}
P_3 =  P_3^*  + b (P_0 - P_0^*) 
\label{linear}
\end{equation}
where $b = 1/(2 - p^\star) = 0.83885634$.   
As a final step, we hypothesize that Eq.\ (\ref{linear}) represents an approximation to $p_c$ of the ``full" kagom\'e system with both up and down triangles shown in Fig.\  \ref{fourfigs}(d).  The justification is that in going from (a) to (b), we replaced one set of stars by shaded triangles satisfying (\ref{linear}), and the system remained at criticality.  Now we replace the second identical set of stars by the same shaded triangles, and we expect that the system remains close to criticality.
\begin{figure}
\includegraphics[width=1\hsize]{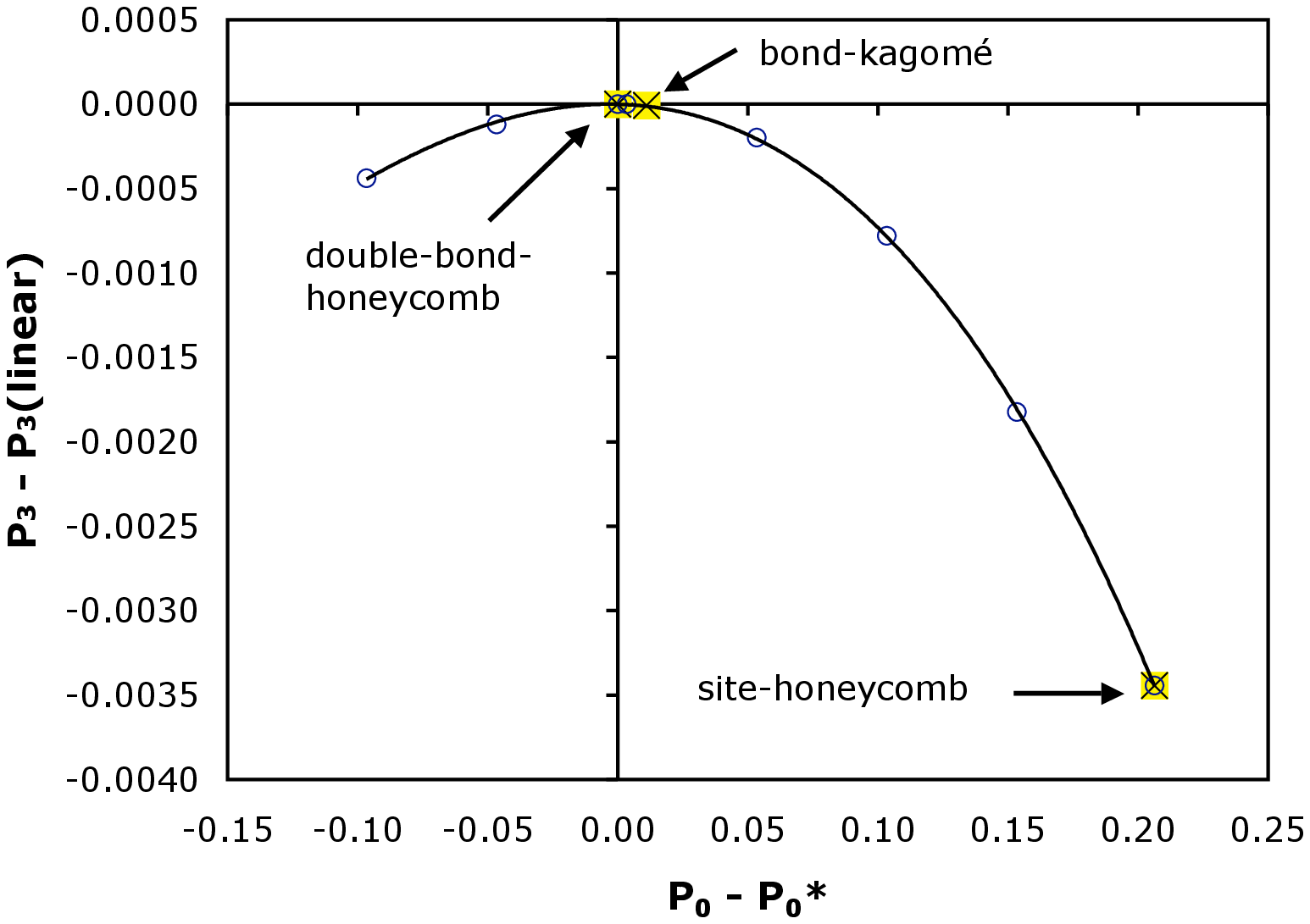}
\caption{Plot of $P_3 - [P_3^* + b (P_0 - P_0^\star)]$ vs.\ $P_0 - P_0^*$ using data of Table \ref{table:P0P2P3}, showing deviations from Eq.\ (\ref{linear}).
Points are numerical data, and the curve is a plot of Eq.\ (\ref{cubic}).  The locations of some specific systems are also shown.}
\label{siriusdifference}
\end{figure}

\begin{table*}
\caption{\label{table:2} Results of $p_c$ and  $P_0$, $P_2$ and $P_3$ 
for various system. $^a$Determined by Eq.\ (\ref{linear}), $^b$determined by Eq.\ (\ref{cubic}), $^c$Ref.\ \cite{ScullardZiff06}, $^d$Ref.\ \cite{HoriKitahara04}, $^e$Ref.\ \cite{Parviainen07}, $^f$Ref.\ \cite{FengDengBlote08}, $^g$this work.   $P_0$, $P_2$ and $P_3$ 
are calculated using $p_c$(cubic).}
\begin{ruledtabular}
\begin{tabular}{lllllll}
system    &      $p_c$(linear)$^a$ &   $p_c$(cubic)$^b$ &       $p_c$(numerical) & $P_0$  & $P_2$ & $P_3$ \\
\hline
double honeycomb &   0.80790076 &   --- &   --- & 0.09652861 & 0.12538387& 0.52731977\\
$(3,12^2)$ & 0.74042118$^c$  & 0.74042081  & 0.74042195(80)$^e$  & 0.10045606   &0.12297685 &  0.53061341\\
kagom\'e  & 0.52440877$^{c,d}$  & 0.52440516  &  0.52440499(2)$^f$   &   0.10757501 & 0.11861544  & 0.53657867  \\
honeycomb (site) &  0.69891402  &  0.69702981  &   0.69704024(4)$^f$ &    0.30297019  & 0 &  0.69702981 \\
\hline
$\infty$ subnet & ---   & ---   &0.628961(2)$^g$ & 0.09652861 & 0.12538387 & 0.52731977\\
subnet 4   & 0.62536437  &  0.62536431 & 0.625365(3)$^g$  & 0.09823481& 0.12433811 & 0.52875085\\
subnet 3   & 0.61933204 &  0.61933180 & 0.6193296(10)$^g$ &0.10016607& 0.12315455 & 0.53037028 \\
subnet 2   & 0.60086322 & 0.60086202 & 0.6008624(10)$^g$ & 0.10402522 & 0.12078995 & 0.53360494\\\end{tabular}
\end{ruledtabular}
\end{table*}

In Table \ref{table:2} we compare the predictions of the linear relation (\ref{linear}) with the numerical results for several systems.  The $p_c$(linear) estimates are found by putting the corresponding expressions for $P_0$ and $P_3$ into Eq.\ (\ref{linear}) and solving numerically for $p$.  For the kagom\'e lattice, we use
\begin{equation}
P_0 = q^3, \qquad 
P_3 = p^3+3p^2 q \ .
\label{kagP}
\end{equation}
For the $(3,12^2)$-lattice (shown for example in Ref.\ \cite{ScullardZiff06}) we use\begin{eqnarray}
P_0 &=&1 - 3 p^2 - 3 p^3 +6 p^{7/2} +3 p^4 -4 p^{9/2} \cr
P_3 &=& 3 p^{7/2} - 2 p^{9/2}  \ .
\end{eqnarray}
For 
site percolation on the honeycomb lattice, $p_c = P_3 = 1 - P_0$, and Eq.\ (\ref{linear}) yields explicitly
$p_c  = 1/[{p^\star}^2(3 - p^\star)] = 0.69891402$.  
The agreement between $p_c$(linear) and numerical results is especially good for systems where $P_0$ is near $P_0^\star$.

To test the behavior of $P_3(P_0)$ over a more complete range of values, we carried out new simulations using the gradient percolation method \cite{RossoGouyetSapoval85,ZiffSapoval86} on a general kagom\'e systems.  We fixed $P_0 = 0, 0.5, 0.1, 0.15,$ and $0.25$  and allowed $P_3$ to vary linearly in the vertical direction, with the estimate of the critical value found as the fraction of $P_3$-triangles in the frontier.  We considered systems of different gradients and extrapolated the estimates to infinity to find the values of $P_3$ given in Table \ref{table:P0P2P3}.

In Fig. \ref{siriusdifference}  we plot the difference between the measured $P_3$ and the predictions of Eq.\ (\ref{linear})
as a function of $P_0$ for these systems.  The first derivative at $P_0 = P_0^*$ appears to be zero, which would imply that Eq.\ (\ref{linear}) represents the \emph {exact} linear term in the behavior of $P_3$ vs.\ $P_0 - P_0^*$.  The numerical data also suggests that (\ref{linear}) gives an upper bound for $P_3(P_0)$ for all $P_0$.    Fitting the data to a cubic equation, assuming that $P_3'(P_0^\star) = b$ exactly, we find
\begin{equation}
P_3 = P_3^* + b (P_0 - P_0^*) + c(P_0 - P_0^*)^2 + d(P_0 - P_0^*)^3
\label{cubic}
\end{equation}
with $c = -0.05987$ and $d = -0.1038$.  This curve fits all the data points $P_3$ within $\pm 10^{-5}$.  The results of using this equation to predict $p_c$ are shown in Table \ref{table:2} under the heading ``cubic", and all are within the expected error of about $\pm10^{-5}$, and more accurate as $P_0$ approaches $P_0^\star$.   For the kagom\'e case, our prediction $p_c = 0.52440516$  compares favorably to the recent precise result  0.52440499(2) of Ref.\ \cite{FengDengBlote08} (which appeared after our analysis was complete) and the previous value 0.5244053(3) \cite{ZiffSuding97}.

\begin{figure}
\includegraphics[width=\hsize]{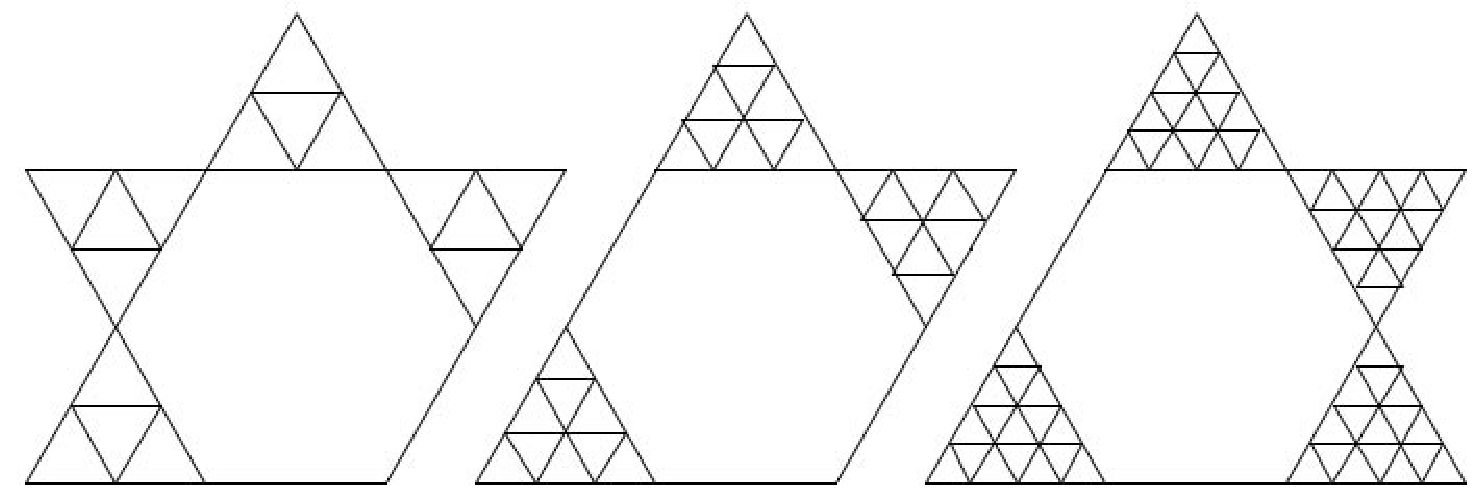}
\caption{Lattices with subnets 2, 3 and 4 (left to right)}
\label{figr18_2}
\end{figure}

We next apply our general  relation for $P_3$ vs.\ $P_0$ to get very accurate $p_c$'s for a class of lattices 
in which each triangle of the kagom\'e arrangement contains a ``stack-of-triangles" as shown in Fig.\ \ref{figr18_2}.
In Ref.\ \cite{HajiakbariZiff08} the similar stack-of-triangles were studied in a regular triangular arrangement, and explicit expressions for $P_0$ and $P_3$ were found by exact enumeration for these three subnets.   We can use those same expressions to analyze the subnets on the kagom\'e lattice as well.
For subnet 2,  we have \cite{HajiakbariZiff08}
\begin{eqnarray}
P_0 &=&q^9+9 p q^8+33 p^2 q^7+54 p^3 q^6+21 p^4 q^5+3 p^5 q^4 \nonumber \\
P_3 &=& 9p^4q^5+57p^5q^4+63p^6q^3+33p^7q^2+9p^8q+p^9  \nonumber \\
\end{eqnarray}
with $q = 1 - p$.
For subnets 3 and 4, see  Ref.\ \cite{HajiakbariZiff08}.

We insert these expressions for $P_0$ and $P_3$ into Eqs.\ (\ref{linear}) and 
(\ref{cubic}) to find the linear and cubic estimates for $p_c$.  The resulting values are shown in Table \ref{table:2}, along with results of numerical simulations.
For subnets 3 and 4, the predictions of (\ref{linear}) and especially (\ref{cubic})  are expected to be very accurate, because $P_0$ is so close to $P_0^*$, and indeed the precision
of the numerical simulations is not high enough to see the difference between these predictions and the actual values.

As seen in Table \ref{table:2}, the quantities $P_0$, $P_2$ and $P_3$ evidently approach the double-honeycomb values $P_0^*$, $P_2^*$ and $P_3^*$ as the mesh of the subnet gets finer.   This is because the triangular units in the fine-mesh limit can be effectively represented by a star of three bonds, with the central site in this star representing the supercritical ``infinite cluster" in the central region of the triangular units \cite{HajiakbariZiff08}.   The set of these stars creates the double-honeycomb lattice, so the $P_i$ are the same as the double-honeycomb values.  Furthermore, the probability $P_{\infty,\mathrm{corner}}$ of connecting from a corner to the central infinite cluster at criticality must be identical to the double-honeycomb bond threshold, $p^\star$.  
Thus, we can find $p_c$ for the infinite net by running simulations of growing clusters from the corner of a single large triangular system, and adjusting $p$ until $P_{\infty,\mathrm {corner}}(p)=p^\star$.   This yields $p_c(\infty) = 0.628961(2)$.

\begin{table}
\caption{\label{table:P0P2P3} Results of simulations for $P_3$ and $P_2 = (1 - P_0 - P_3)/3$ for
general  kagom\'e systems as a function of $P_0$; values are accurate to about $10^{-6}$.  These
data are plotted in Fig.\ \ref{siriusdifference}.  Also shown are the equivalent site-bond
probabilities $p_s$ and $p_b$ calculated from Eqs.\ (\ref{pands}).  The third row is the double-honeycomb system and the final row represents site percolation on the honeycomb lattice \cite{FengDengBlote08}.}
\begin{ruledtabular}
\begin{tabular}{lllll}
$P_0$  & $P_2$ & $P_3$ & $p_b$ & $p_s$ \\
\hline
0  &	0.1846972  &	0.4459084  & & \\
0.05  &	0.1539432  &	0.4881704  & & \\
0.0965286 & 0.1253839& 0.5273198  & 0.6527036	&	1 \\
0.1  &	0.1232560  &	0.5302320  & 0.6583497	&	0.9926153 \\
0.15  &	0.0926739  &	0.5719784  & 0.7405771	&	0.8974788 \\
0.2  &	0.0622208  &	0.6133375 & 0.8242773	&	0.8195766 \\
0.25  &	0.0319205  &	0.6542385 &  0.9091230	&	0.7547482 \\
0.3029598 &	0  &	0.6970402  &1 & 0.6970402 \\
\end{tabular}
\end{ruledtabular}
\end{table}

Finally, we note that a realization of the general kagom\'e system for $P_0 \ge P_0^*$ is given by site-bond percolation on the honeycomb lattice, as represented in 
Fig.\ \ref{honeycombfig}.   For the site-bond basic unit  of Fig.\ \ref{sitebondfig}, we have
\begin{eqnarray}
P_0 &=& 1-p_s + p_s [(1-\sqrt{p_b})^3 + 3 (1-\sqrt{p_b})^2 \sqrt{p_b}] \nonumber \\
P_3 &=& p_s  p_b^{3/2}
\label{sitebondeq}
\end{eqnarray}
which can be inverted to yield:
\begin{equation}
p_b = \left(\frac{3 P_3}{2 P_3 - P_0 + 1} \right)^2  \ , \qquad p_s = P_3/p_b^{3/2}
\label{pands}
\end{equation}
In Table \ref{table:P0P2P3}, we list the values of $p_b$ and $p_s$ that correspond to the 
measured values of $P_3(P_0)$.  
We can also put Eq.\ (\ref{sitebondeq}) into Eq.\ (\ref{linear}) and simplify using Eqs.\ 
(\ref{P0star}) and (\ref{P3star}) to find an  approximate expression for the critical line on the $p_s$--$p_b$ plane:
\begin{equation}
p_s = \frac{{p^\star}^2}{p_b( 1 - B (\sqrt{p_b} - p^\star))}
\end{equation}
where $B = p^\star/(3 - {p^\star}^2)$.  
We can improve upon this relation by using
the cubic function of $P_3(P_0)$ given in Eq.\ (\ref{cubic}); this adds the additional terms $C (\sqrt{p_b} - p^\star)^2 + D (\sqrt{p_b} - p^\star)^3$ to the above formula, where $C = 9 {p^\star}^2(2 - p^\star)^3/(3 - {p^\star}^2)^3 c = -0.0460682$ and $D = -0.01681$.

In conclusion, we have shown how the notion of a unique relation between $P_3$ and $P_0$, first studied in the context of self-dual systems \cite{Ziff06,ChayesLei06}, extends to the non-self-dual kagom\'e configuration.  The approximate linear expression we found, Eq.\ (\ref{linear}), appears to be exact to first order, and the simulation results shown in Fig.\ \ref{siriusdifference} suggest that that expression provides
upper bounds to $p_c$ for these systems.  We conjecture that this is indeed the case.  The numerically refined cubic relation of Eq.\ \ref{cubic} allows very accurate thresholds to be predicted for a wide variety of systems, and an explicit expression for the criticality condition of site-bond percolation on the honeycomb lattice to be written.

This work was supported in part by the U. S. National Science Foundation Grant No.\ DMS-0553487.

\bibliographystyle{apsrev}
\bibliography{SiriusZiffRapid}


\end{document}